\documentclass[a4paper,fleqn]{cas-dc}
\usepackage{hyperref}
\usepackage[authoryear]{natbib}
\usepackage{graphics}
\usepackage{graphicx}
\usepackage{subcaption}
\usepackage{caption}
\usepackage[switch]{lineno}

\begin{document}
\let\WriteBookmarks\relax
\shorttitle{Cutting device attitude control of cabbage harvester}
\shortauthors{Y. Park, H. I. Son}

\title [mode = title]{A Sensor Fusion-based Cutting Device Attitude Control to Improve the Accuracy of Korean Cabbage Harvesting} 
\tnotetext[1]{This research was supported by the Korea Institute of Planning and Evaluation for Technology in Food, Agriculture and Forestry(IPET) through the Agriculture, Food and Rural Affairs Convergence Technologies Program for Educating Creative Global Leader Program, funded by the Ministry of Agriculture, Food and Rural Affairs(MAFRA)(716001-7)}
\author[1]{Yonghyun Park}[orcid=0000-0001-8904-4234]
\ead{dk03378@nate.com}
\credit{Methodology, hardware, Software, Validation, Writing-Original Draft, Visualization}

\author[1,2]{Hyoung Il Son}[orcid=0000-0002-7249-907X]
\ead{hison@jnu.ac.kr}
\credit{Conceptualization, Writing-Review \& Editing, Supervision, Project administration, Funding acquisition}

\cormark[1]
\ead[URL]{www.hralab.com}

\address[1]{Department of Rural and Biosystems Engineering, Chonnam National University, Yongbong-ro 77, Gwangju 61186, Republic of Korea}
\address[2]{Interdisciplinary Program in IT-Bio Convergence System, Chonnam National University, Yongbong-ro 77, Gwangju 61186, Republic of Korea}

\cortext[cor1]{Corresponding author. Yongbong-ro 77, Gwangju 61186, Republic of Korea}
\begin{abstract}
Korean cabbage harvesting lacks mechanization and depends on human power; thus, conducting research on Korean cabbage harvesters is of immense importance. Although these harvesters have been developed in various forms, they have not yet attained commercialization. Most Korean cabbage fields have slopes; thus there are several challenges, that can prevent accurate harvesting. Therefore, to address these challenges at the site, we adopt two cylinders in this study, develop a mechanism that enables attitude control of the cutting device, not driving platform body, to cope with slopes. By maintaining the level, angle, height of cutting, we can reduce loss and improve harvest performance. It is difficult to find examples where these mechanisms have been applied. For basic research, sensor fusion has been carried out based on the Kalman filter, which is commonly utilized for attitude control. The hydraulic cylinder was controlled using the data obtained for maintaining the attitude. Furthermore, field tests were conducted to validate this system, and the root mean square error (RMSE)  was obtained and verified to quantitatively assess the presence or absence of attitude control. Therefore, the purpose of this study is to suggest a development direction for Korean cabbage harvesters via the proposed attitude control system.
\end{abstract}

\maketitle { }
\section{Introduction}
Korean cabbage, which is a vegetable with high consumption and high production rates, is the main ingredient of Kimchi, one of Korea's major foods. However, owing to the lack of sufficient mechanization and automation for the harvest process, which requires considerable amounts of time and effort. This vegetable is almost entirely cultivated by human power. Therefore, research and development are required to mechanize the harvest process of cabbage. Currently, several cabbage harvesters with similar varieties have been developed and commercialized globally. A research on the development of a Korean cabbage harvester was conducted in a manner similar to a cabbage harvester, and a development study was conducted in the form of walking and pull-type harvesters ~\citep*{cao2020design, el2020fabrication}. However, unlike cabbages whose stems and roots are outside their head regions, the roots and stems of Korean cabbages have geometric shapes inside their head regions. In addition, owing to the limitations of the cultivation method and field depth, development of Korean cabbage harvesters is challenging\citep{ali2019kinematic}. As the body of the Korean cabbage harvester is inclined, it alters the level and cutting height of the cutting device, thereby preventing accurate cutting and impeding harvest performance. Depending on the cutting attitude of the cutting device, Korean cabbage harvesters experience challenges such as missed cuts, over cuts, and side cuts. If the roots and stems are missed cut, additional machining operations are required, and in the case that they are side cuts or over cuts, quality concerns are inevitable owing to the damage to the head region of the Korean cabbage. Consequently, accurate attitude control is required.

The attitude control of agricultural machinery has been developed by research on devices that maintain their level by controlling attitudes, such as the orbital combine and tractor three-point linkage (\cite*{kise2006sensor}; \cite{sun2020design}). However, this device levels the body of the harvester with the ground, and is limited in its ability to accurately control the position of the cutting device. In addition, although sensor installation is required to measure attitude data, indirect sensors, such as lasers and ultrasound, are disadvantageous owing to the dust, mud, water, and outer leaves of the crop. Therefore, it is necessary to study and develop a novel  mechanism for correct attitude control, as described earlier, as well as attach direct sensors suitable for agricultural machinery.

Studies on attitude control are being actively conducted in various fields, such as in the fields of automobiles, drones \citep*{guo2017novel}, mobile robots\citep*{odry2018kalman}, telemetry and satellites\citep*{du2011finite, kim2019tracking}. Substantial efforts have been made to control attitude accurately with the required attitude adopting various sensors(\cite{zhu2019multi}; \cite*{ jiang2016fixed, zhou2020data}). To address the limitation  of a single sensor, several cases of sensor fusion adopt a Kalman filter, which fuses multiple sensors to estimate accurate data\citep*{konigseder2016attitude}. However, in the agricultural sector, critical attention is paid to platform driving, drone control, and robots, whereas the studies conducted in other fields are insufficient. In particular, it has rarely been applied to harvesting operations that lack mechanization and automation, and most of these operations are controlled by a single sensor. Therefore, it is necessary to expand the research direction for accurate harvesting.

In this study, we devised a mechanism for controlling the attitude of the cutting device of a Korean cabbage harvester that requires accurate cutting and, presented the basis for developing automatic harvesters. This mechanism can solely control the cutting device; therefore, that it is unaffected by alterations in the harvester’s body attitude. Because the device is coupled to the underside of the cutting device, which makes it simpler to structure than body attitude control, it can also be applied to other driving platforms, thus increasing its general application. The proposed mechanism comprises pitch, slide, and roll shafts for maintaining the cutting angle of the cutting device, maintaining the cutting height, and ensuring level control, respectively. In addition, it is designed by introducing a cutting machine guide mechanism that can measure the cutting height using a linear potentiometer. These mechanisms have no applied case; therefore, for the basic research, we employed a Kalman filter with a few applications in agriculture to design the control algorithms. The algorithm fused the accelerometers, gyro sensors, and linear potentiometers to estimate the attitude using Kalman filters. Furthermore,  proportional–integral–derivative (PID) control was integrated for stable attitude control in response to rapid  attitude alterations in slopes or obstacles. In addition, field tests were conducted to verify the proposed attitude control system, and the attitude of the cutting device  was  investigated and evaluated with and without attitude control.
\section{Previous Harvester}
\subsection{Structure}
\label{sec2.1}
\begin{figure}[!t]
	\centering
		\includegraphics[width=0.76\linewidth,page=0]{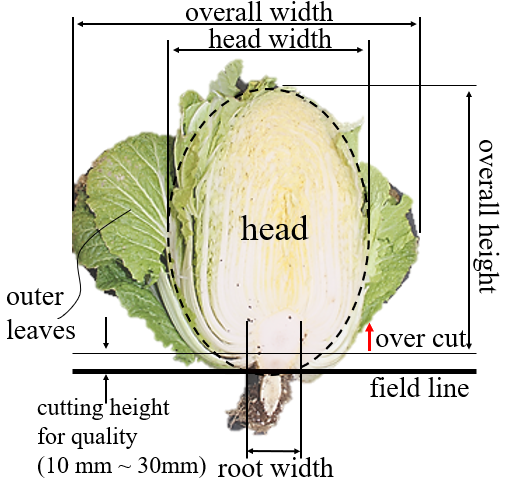}
	\caption{Korean cabbage structure(Section).}
	\label{f1}
\end{figure}

\begin{figure} 
  \centering
  \begin{subfigure}[b]{.78\linewidth}
    \centering
    \includegraphics[width=\linewidth]{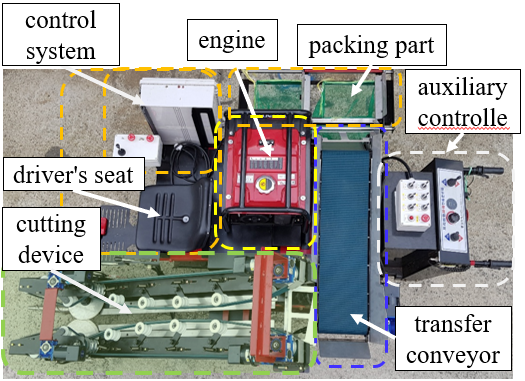}
    \caption{}
  \end{subfigure}
  \begin{subfigure}[b]{.78\linewidth}
    \centering
    \includegraphics[width=\linewidth]{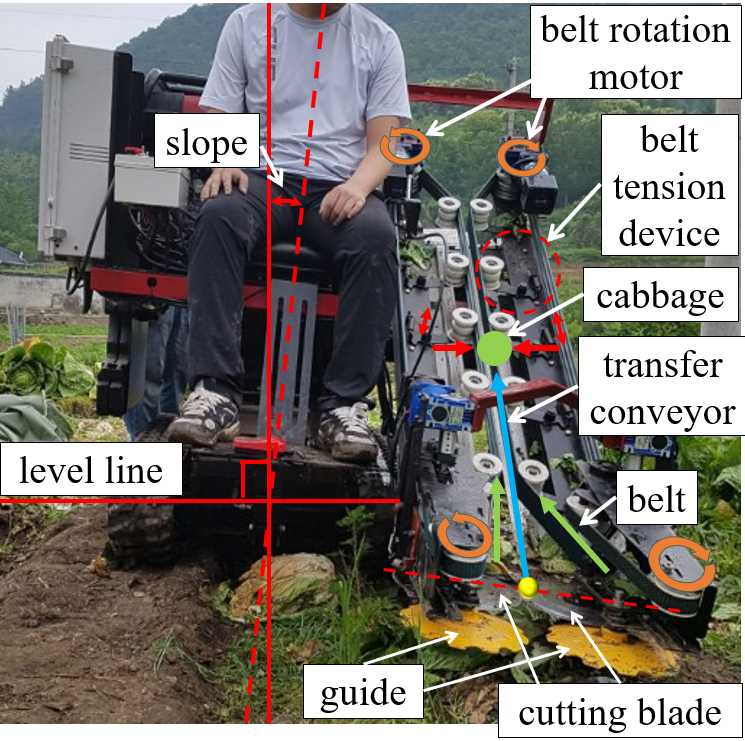}
    \caption{}
  \end{subfigure}\\
\caption{The structure of the previous Korean cabbage harvester; (a) Top view, (b) Front view.}
\label{f2}
\end{figure}

As illustrated in Fig. \ref{f1}, Korean cabbages have geometric shapes with stems and roots inside the conjunctiva, unlike other cabbages. The height of the cut determines the amount of outer leaves removed, and inaccurate cuts can cause damage to the head and affect quality. Accordingly, a harvesting machine was developed to accurately cut Korean cabbage during harvests, as illustrated in Fig. \ref{f2}. This machine has a lift-type height control unit for adjusting the cutting height, this mechanism allows the cutting height to be adjusted manually. Therefore, it increases the possibility of a more accurate harvest. As a detailed description,when cutting Korean cabbages with the cutting blade presented in Fig. \ref{f2}(b), the cutting device moves the cabbages through the transfer belt to the transfer conveyor. The transfer part comprises several springs and conveyor belts. The conveyor belts on both sides maintain tension owing to their optimal spring elasticity. A press force is also exerted on the Korean cabbage in contact with the conveyor belt from the side. The motor rotates the conveyor belt, while the cabbages are moved by the lateral forces maintained by the spring configured in succession. Then, the cabbages moved from the transport conveyor to the packaging work part were packaged, and the harvest was completed.

\begin{figure} 
  \centering
  \begin{subfigure}[b]{.42\linewidth}
    \centering
    \includegraphics[width=\linewidth]{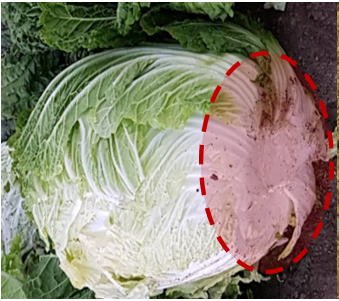}
    \caption{}
  \end{subfigure}
  \begin{subfigure}[b]{.42\linewidth}
    \centering
    \includegraphics[width=\linewidth]{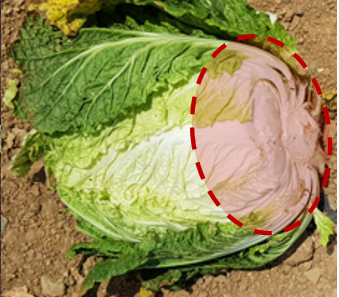}
    \caption{}
  \end{subfigure}\\
  \begin{subfigure}[b]{.42\linewidth}
    \centering
    \includegraphics[width=\linewidth]{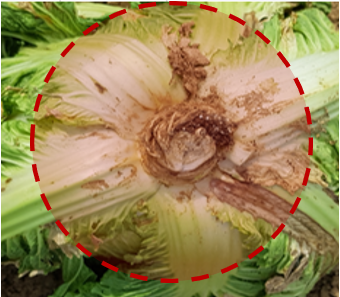}
    \caption{}
  \end{subfigure}
  \begin{subfigure}[b]{.42\linewidth}
    \centering
    \includegraphics[width=\linewidth]{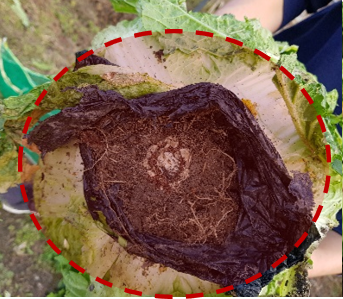}
    \caption{}
  \end{subfigure}
\caption{Challenges encountered during amputation; (a) Over cutting, (b) Side cutting, (c) Missed cutting, (d) Vinyl cutoff.}
\label{f5}
\end{figure}
\subsection{Limitations}
\label{sec2.2}
\begin{figure}[!t] 
    \centering
    \includegraphics[width=1.0\linewidth,page=1]{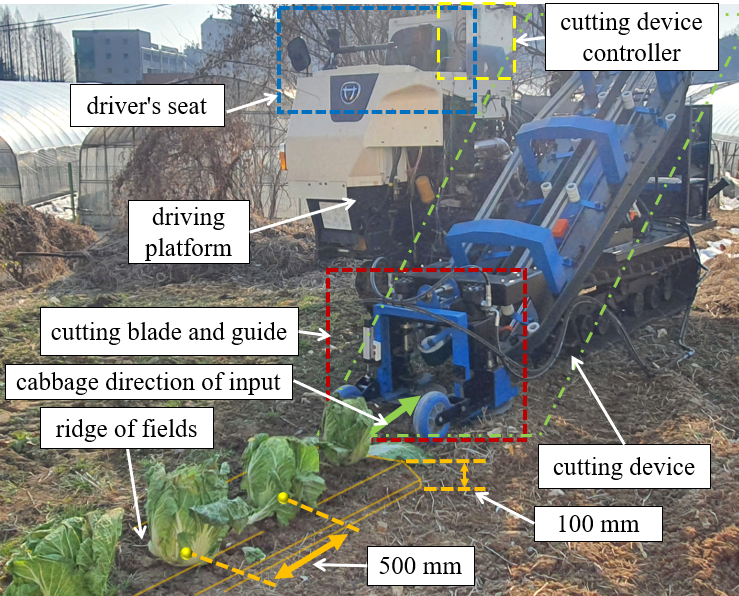}
    \caption{Developed Korean cabbage harvester}
    \label{f6}
\end{figure}

As illustrated in Fig. \ref{f2}(b), the inclination of the harvester’s body impedes accurate cutting during harvesting. Accordingly, as illustrated in Fig. \ref{f5}, challenges such as over cut (Fig. \ref{f5}(a)), and side cut of cabbage with diagonal sides trigger poor quality Korean cabbages (Fig. \ref{f5}(b)). Some farms use vinyl mulching. therefore, when cutting, vinyl was cut off(Fig. \ref{f5}(c)), missed cut roots (Fig. \ref{f5}(d)). Furthermore, in Korea, where there are many mountainous areas, most Korean cabbage fields are cultivated on sloping land. Therefore, the limitation of the body slope of the harvester is more prominent; hence the need to develop Korean cabbage harvesters that can address the demand of field slopes and the obstacles accompanying them. To address these challenges, this study  proposes a mechanism that solely enables attitude control on the cutting device, as illustrated in Fig. \ref{f6}.
\subsection{Attitude Control Mechanism of Cutting Device}
\label{sec2.3}
Korean cabbages always grow vertically toward the sky while facing the sun. However, the field conditions of cultivated fields differ significantly. Field conditions trigger inclinations and height differences relative to the target. For accurate cutting, it is necessary to maintain a specific attitude and position on the system. Accordingly, we define states that require control as follows: 

\begin{itemize}

\item Left and right roll angle ($\phi$) : \\Side direction, maintain level control
\item Up and down pitch angle ($\theta$): \\Travel direction, maintain cutting device angle
\item Height difference from Cutting point ($H_{p}$): \\Retain cutting point of cutting device
\end{itemize}
We also specify the mechanisms for controlling and compensating each axis, as illustrated in Fig. \ref{f7}(a). 
\subsubsection{Maintaining level attitude for side direction($\phi$)}
In general, during the preparations for cultivating Korean cabbages, ridges are created following the growth direction. Therefore, even if the ground is tilted from side to side, the cutting part is required to be continuously horizontal. Consequently, we apply a mechanical structure in which $\phi$ can move as the blade part of the cutting device, operates alone, and can be autonomously compensated by gravity when driving while grounding the field. The additional two connected wheels allow natural movements along the ridge.
\subsubsection{Maintaining cutting attitude for travel direction ($\theta$)}
In contrast to the side direction, it is challenging to maintain an attitude toward the travel direction. This is because the terrain of the cultivated land is very different, and the terrain is primarily influenced by its geographical features. In addition, the shape of the soil in the travel direction changes with the weight of the agricultural machinery. The control, as presented in Fig. \ref{f7}(b), is adopted to control the axis of $\theta$ by operating \textcircled{\small 1} cylinder stroke $L_{\theta c}$ to maintain the control objective $\theta_{c}$ of the angle between the ground and the cutting device. Owing to this device, the cutting device operates while maintaining the control objective $\theta_{c}$, regardless of the angle of the driving platform body.
\subsubsection{Maintaining the cutting-point height ($H_{p}$)}
\begin{figure*} 
  \centering
  \begin{subfigure}[b]{.28\linewidth}
    \centering
    \includegraphics[width=\linewidth]{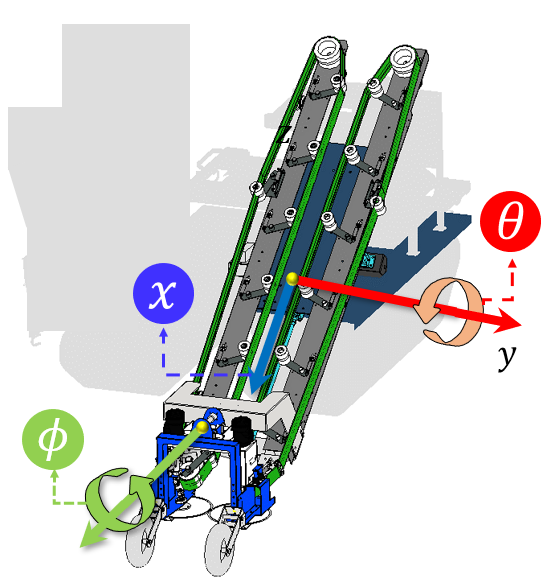}
    \caption{}
  \end{subfigure}
  \begin{subfigure}[b]{.35\linewidth}
    \centering
    \includegraphics[width=\linewidth]{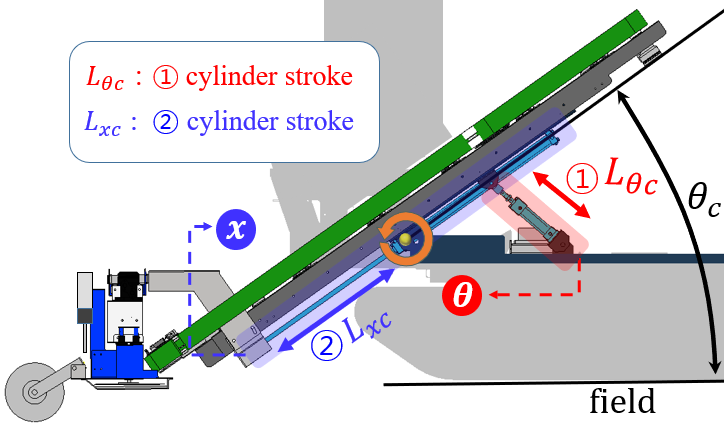}
    \caption{}
  \end{subfigure}
  \begin{subfigure}[b]{.35\linewidth}
    \centering
    \includegraphics[width=\linewidth]{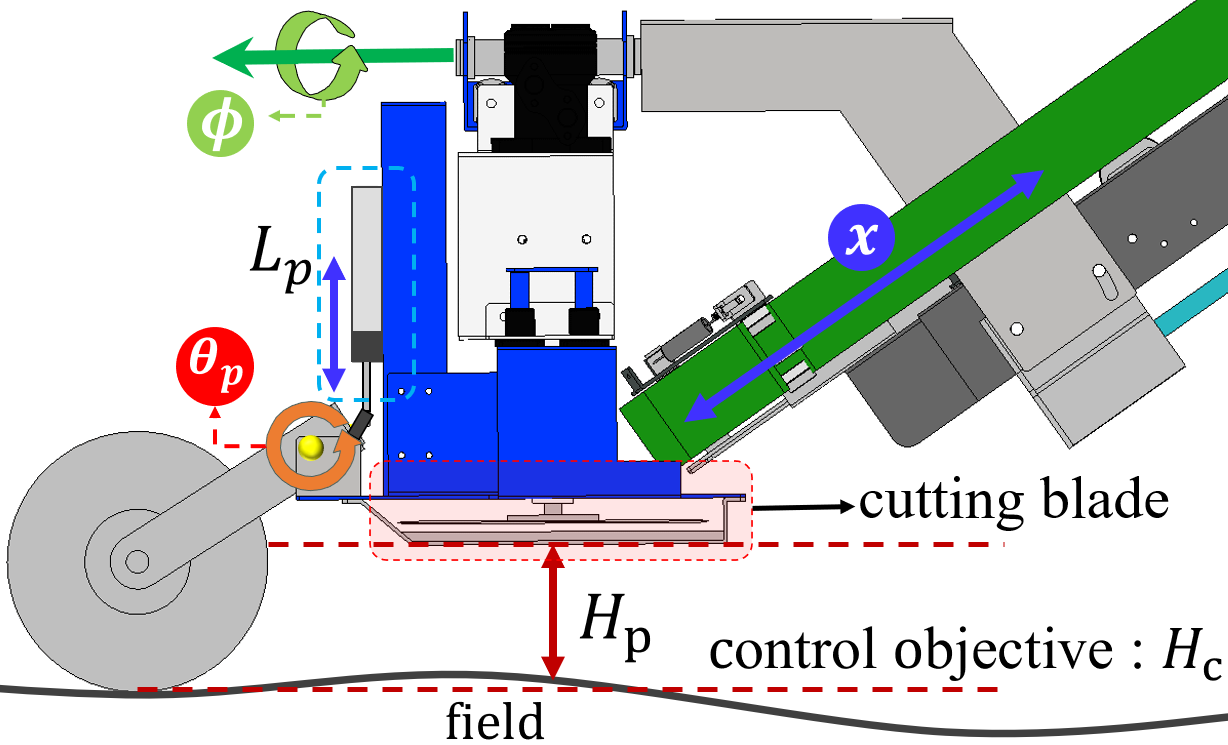}
    \caption{}
  \end{subfigure}
\caption{Cutter attitude control mechanism; (a) Stereoscopic, (b) Right side view, (c) $H_{p}$ height measurement guide mechanism.}
\label{f7}
\end{figure*}

Even if the optimal cutting angle is maintained, the cut cabbage loses its value as a product provided the cutting height is not suitable. $\theta_{p}$ is used to maintain a proper height to the ground, and is obtained from wheels in contact with the ridge. The wheel has a structure that moves up and down owing to the contact forces from the field. The measured values were calibrated with the height $H_{p}$ between the ground and the cutting blade for optimal height, thus allowing the data to retain the control objective $H_{c}$. The height control slides the $x$ axis by driving the \textcircled{\small 2} cylinder stroke $L_{x c}$ in Fig. \ref{f7}(b). The up and down slopes for the attitude control and cutting-point height alterations are closely related. If either of the two is not suitably controlled, the Korean cabbage is not ideally cut. Therefore, the operation continuously inspects each condition to maintain a suitable cutting attitude.
\section{Sensor Fusion-based Algorithm}
Applications in agriculture have been studied using the Kalman filter to estimate the location of unmanned agricultural vehicles between fruit trees and crops, measure systems to control drone attitude, and improve robot work performance
(\cite*{konigseder2016attitude}; \cite{blok2019robot, ball2016vision}; \cite*{kim2019unmanned}). However, more studies are required on the harvesting process in which the attitude of the harvesters determines the crop quality. The cutting device attitude control mechanism proposed in this study was not applied at home and abroad, and the basic research was carried out by conducting sensor fusion with a Kalman filter, which has fewer applications in the agricultural sector.
\subsection{Kalman Filter-based Sensor Fusion}  
\begin{figure*}
	\centering
	\begin{subfigure}[!t]{.9\linewidth}
    \centering
    \includegraphics[width=\linewidth]{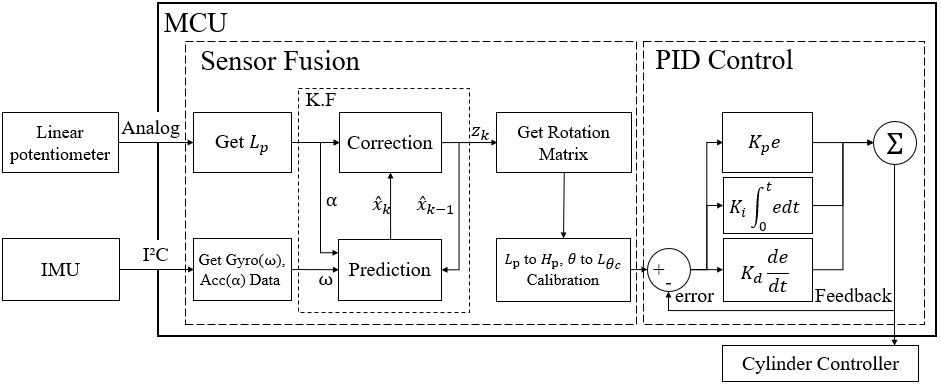}
  \end{subfigure}
	\caption{Attitude control system algorithm.}
	\label{f8}
\end{figure*}

We designed an algorithm for controlling the cutting device attitude, as presented in Fig. \ref{f8}. The gyro sensor has angular velocity $\omega$ and accelerometer acceleration $\alpha$, and it fused the analog data $L_p$ obtained from the linear potentiometer into the Kalman filter. The output rotational $\theta$ and length $L_p$ matrices maintain the attitude of the cutting device by controlling the cylinder with the data calibrated using the cylinder length presented in Fig. \ref{f8}. The slide($x$) axis \textcircled{\small 2} cylinder is controlled, such that the $L_p$ of the length matrix can be calibrated with the height $H_{p}$ between the ground and the cutter blade to maintain the required height and enhance the stability of attitude control.
\subsection{Attitude Control}
For Kalman filter system modeling, we referred to the algorithm presented in\citep*{odry2018kalman}.
The acceleration sensor, which is slow to reach the actual value, but returned to zero accurately from a standstill, is calibrated with a gyro sensor that exhibits a severe error accumulation but is strong at reading patterns.

First, even if the angular velocity measured by the gyro sensor is immediately integral, the Euler angle cannot be obtained. Therefore, the measured value of the gyro sensor should be changed to the rate of change in the Euler angle.
The relationship between Euler angles $[\phi,\theta,\psi]$ and angular velocity $[p,q,r]$is expressed as
\begin{equation}
\begin{bmatrix}\dot{\phi} \\ \dot{\theta} \\ \dot{\psi} \end{bmatrix} = \begin{bmatrix}
1  & \sin\phi\tan\theta & \cos\phi\tan\theta \\
0  & \cos\phi & -\sin\phi \\ 
0  & \sin\phi/\cos\theta & \cos\phi/\cos\theta
\end{bmatrix}\begin{bmatrix}p \\ q \\ r \end{bmatrix}.
\end{equation}
After determining the Euler angle, it was inferred to be integral. We then obtained an angle $\theta$, a pitch angle, as defined by (2),
\begin{equation}\label{eq1}
\theta_{cur} = \theta_{pre} + \Delta t(q\cos\phi-r\sin\phi).
\end{equation}
The value of $\theta_{cur}$ becomes $\theta$ of the gyro sensor. Next, the formula (3) for changing the acceleration$[X, Y, Z]_{acc}$ measured by the accelerometer to the Euler angle is as follows,
\begin{equation}\label{eq2}
\theta_{acc} = \arctan(Y_{acc} / Z_{acc}) \times 180 / \pi.
\end{equation} 
The calculated Euler angle was applied to the Kalman filter algorithm. First, the state equation of the system for time $k$ is expressed as
\begin{equation}
x_{k}=Ax_{k-1}+Bu_{k}+Q_{k}.\qquad\
\end{equation} 
where the state variable matrix is $x_{k}=[\theta, \dot{\theta}_{b}, L_{p}]^{T}_{k}$, where $\dot{\theta}_{b}$ is the amount in which the angle $\theta$ is accumulated over time, which is called bias. It reduces the error of the measured value by subtracting the bias from the measured value of the gyro sensor. The current angle was then obtained as follows
Thus, the transition matrix $A$ from the previous state $x_{k-1}$ is
\begin{equation}
A=\begin{bmatrix}
1 & -\Delta t & 0\\
0 & 1 & 0\\
0 & 0 & 1
\end{bmatrix}.\qquad
\end{equation}
$u$ represents a control input with units of [deg/s], and the gyro sensor measurements at point $k$ and can be written by changing it to $\dot{\theta}$.
where $B$ represents the control input matrix as follows
\begin{equation}
B=[\Delta t, 0, 0]^{T}\qquad\\
\end{equation}
$Q_{k}$ is the covariance matrix of process noise $w_{k}$. This represents the covariance matrix of the state estimates of the accelerometer and the bias. We assumed that the estimations of the bias and accelerometers are independent of each other, the follows 
\begin{equation}
Q_{k}=\begin{bmatrix}
Q_{i}\Delta t & 0                    & 0\\
0                    & Q_{i}\Delta t & 0\\
0                    &    0                  & Q_{p}\Delta t
\end{bmatrix}.
\end{equation}
As in the expression above, the covariance matrix $Q_k$ depends on the current time $k$. To obtain the observation model $H$, the follows
\begin{equation}
z_{k}=H\hat{x}_{\bar{k}}+v_{k}.\quad
\end{equation}
As in the expression above, $H$ is a constant matrix and represents the relationship between the measurements and the state variables, the follows 
\begin{equation}
H=\begin{bmatrix}
1 & 0 & 0\\
0 & 0 & 1
\end{bmatrix}.
\end{equation}
$R$ is Matrix representing measurement noise. Thus, the expression $R$ offsetting only the noise of $\theta_{acc}$ of accelerometers and $L_p$ of linear potentiometers is as follows
\begin{equation}
R=
\begin{bmatrix}
R_{i}& 0 \\
0 & R_{p}
\end{bmatrix}.\;
\end{equation}
All system models were obtained. Equations (5), (6) (7), (9), and (10) are substituted to the Kalman filter algorithm to obtain $\theta$ and $L_{p}$ of Fig. \ref{f8}.
When Kalman filter algorithms are applied, as the following expression is obtained
\begin{equation}
\hat{x}_{\bar{k}}=A\hat{x}_{k-1}+B\dot\theta_{k}.
\end{equation}
When substituting Status variables $x=[\theta, \dot\theta_{b}, L_{p}]$, System matrix
$A$, and Control input matrix $B$ of (5), (6), and (7), respectively, into (11), the equation obtained is expressed as
\begin{equation}
\begin{bmatrix}
\theta\\
\dot\theta_{b}\\
L_{p} 
\end{bmatrix}_{k}=\begin{bmatrix}
\theta+\Delta t(\dot\theta-\dot\theta_{b}) \\
\dot\theta_{b}\\
L_{p} 
\end{bmatrix}_{k}.
\end{equation}
Because we cannot measure bias directly, we adopted a previous method to estimate bias.
The error covariance formula is expressed as 
\begin{equation}
P_{\bar k} = AP_{k-1}A^{T}+ Q_{k}.
\end{equation}
If (5) and (7) are replaced by (13), the error covariance $P_{\bar{k}}$ shall be interpreted as follows
\begin{equation}\begin{aligned}
&\ \ \ \ \ \ \ P_{\bar{k}} =
\\&\footnotesize{\begin{bmatrix}
P_{00}+\Delta t(\Delta t P_{11}-P_{01}-P_{10}+Q_{i})& P_{01}- \Delta t P_{11}& P_{02}- \triangle t P_{12}\\
P_{10}- \Delta t P_{11} & P_{11} +\Delta t Q_{i} & P_{12}\\
P_{20}- \Delta t P_{21} & P_{21} & P_{22} +\Delta t Q_{p}
\end{bmatrix}}. \end{aligned}
\end{equation}
The measurements obtained are expressed (8), and are modified to obtain $v_{k}$.
\begin{equation}
v_{k}=z_{k}-H\hat{x}_{\bar{k}}.\quad
\end{equation}
The angle is adopted as it is because the current state variable on the right is not calibrated. $z_{k}$ is a novel measurement obtain.
$z_{k}$ = $[\theta_{acc}, L_{p}]^{T}_{k}$, where $\theta$ represents the Euler angle measured by the accelerometer. $L_{p}$ is also a new linear potentiometer length, the follows
\begin{equation}
v_{k}=
\begin{bmatrix}
\theta_{acc}  \\
L_{p}  
\end{bmatrix}_{k}-
\begin{bmatrix}
\theta  \\
L_{p}  
\end{bmatrix}_{k-1}.
\end{equation}
The formula for saving the Kalman gain $K_k$ is expressed as 
\begin{equation}
K_{k}=\frac{P_{\bar k}H^{T}}{HP_{\bar k}H^{T}+R}.
\end{equation}
The reliability of the measurements can be predicted using the previous error covariance matrix $P_{\bar{k}}$ and the measurement covariance matrix $R$. The observation model was adopted to map the previous error covariance matrix $P_{\bar{k}}$ to the observation space. The Kalman gain formula can be transformed as follows
\begin{equation}
K_{k} = P_{\bar k}H^{T}S_{k}^{-1}\quad\
\end{equation}
\begin{equation}
S_{k}= HP_{\bar k}H^{T} + R.
\end{equation}
The larger the $S$ value, the more reliable the prediction, and the smaller the value, the more reliable the measurement. If (9) and (10) are substituted, then they are expressed as
\begin{equation}\begin{aligned}
S_{k}=\begin{bmatrix}
P_{00} & P_{02} \\
P_{20} & P_{22} 
\end{bmatrix}_{k}+
\begin{bmatrix}
R_{i}& 0 \\
0 & R_{p}
\end{bmatrix}.\qquad
\end{aligned}
\end{equation}
The size of $R_{k}$ influences the Kalman gain, and it is assumed that the measurement noise is identical and does not depend on time $k=[K_{0},K_{1},K_{2}]^{T}$. By substituting (18), we obtained Kalmangein $K_{k}$, the follows
\begin{equation}
\begin{bmatrix}
K_{0} \\
K_{1} \\
K_{2}
\end{bmatrix}=
\begin{bmatrix}
P_{00}/S & P_{02}/S \\
P_{10}/S & P_{12}/S \\
P_{20}/S & P_{22}/S
\end{bmatrix}_{k}.
\end{equation}
The formula for the estimates is presented below. Replace the $v_{k}$ obtained from (16),
\begin{equation}
\hat{x}_{k}=\hat{x}_{\bar{k}}+K_{k}v_{k}\quad
\end{equation}
\begin{equation}
\begin{bmatrix}
\theta \\
\dot\theta_{b}\\
L_{p}
\end{bmatrix}_{k+1}=
\begin{bmatrix}
\theta \\
\dot\theta_{b}\\
L_{p}
\end{bmatrix}_{k}+
\begin{bmatrix}
K_{0}v_{k} \\
K_{1}v_{k}\\
K_{2}v_{k}
\end{bmatrix}.\
\end{equation}
The error covariance formula is expressed as
\begin{equation}\
P_{k}=(I-K_{k}H)P_{\bar k}.
\end{equation}
Substituting (9) and (18), as well as $P_{k}$ into (24), the solution obtained is expressed as
\begin{equation}\begin{aligned}
P_{k}=\begin{bmatrix}
P_{00} & P_{01} & P_{02}\\
P_{10} & P_{11} & P_{12}\\
P_{20} & P_{21} & P_{22}
\end{bmatrix}_{k-1}-
\begin{bmatrix}
K_{0}P_{00} & K_{0}P_{01} & K_{0}P_{02}\\
K_{1}P_{10} & K_{1}P_{11} & K_{1}P_{12}\\
K_{2}P_{20} & K_{2}P_{21} & K_{2}P_{22}
\end{bmatrix}.\end{aligned}
\end{equation}
$Q$ and $R$ of the Kalman filter system model can be adjusted according to the environment, and selected as experiments to adjust them.

To stabilize the attitude of agricultural machinery, which is often driven in uneven places, PID control systems are integrated into algorithms. By inputting the measured data, overshoots exceeding the control value through proportional, integral, and differential processes are calibrated with PID control to stabilize them. The PID control system is illustrated in the Fig. \ref{f8}. We obtained $[K_{p}, K_{i}, K_{d}]$ for control, and set it to gain [0.1, 0 , 3] for $L_{\theta c}$, and gain [0.1, 0.02, 1] for $H_{p}$.
\begin{figure}[!t] 
    \centering
    \begin{subfigure}[!t]{1.0\linewidth}
        \centering
        \includegraphics[width=\linewidth]{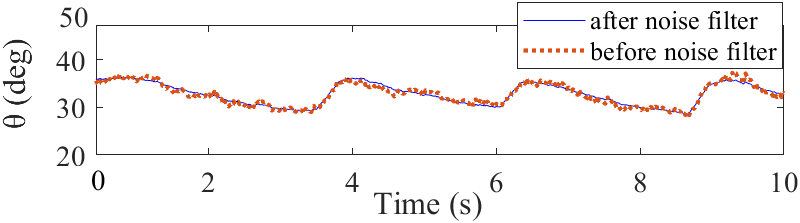}
        \caption{}
    \end{subfigure}
    \begin{subfigure}[!t]{1.0\linewidth}
        \centering
        \includegraphics[width=\linewidth]{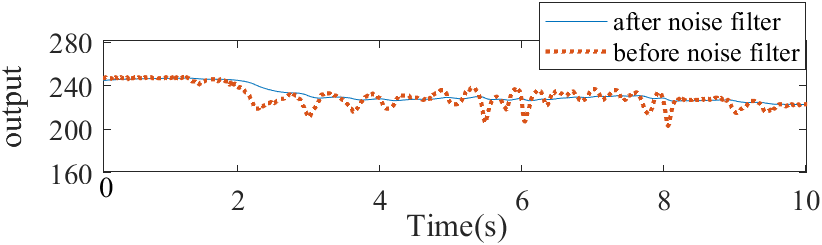}
        \caption{}
    \end{subfigure}
\caption{Sensor data noise reduction; (a) IMU sensor data, (b) Linear potentiometer data.}
\label{f9}
\end{figure}
\section{Korean Cabbage Harvester Field Test}
\subsection{Adjustment of System Model Parameters}

In this study, the performance of the system was evaluated by applying the Kalman filter-based sensor fusion algorithm to the cutting device attitude control mechanism of an actual Korean cabbage harvester. In addition, noise was eliminated from the vibrations of the combined platform by modifying the Kalman filter system variables, Q and R. As expressed in (\ref{eq4A1}), Q and R were selected, and Fig. \ref{f9}

Fig. \ref{f9}(a) presents the noise attenuation results obtained from the gyro and acceleration sensors, while Fig. \ref{f9}(b) represents the noise attenuation results from the linear potentiometer. Using the data obtained from this adjustment, we applied it to the algorithm and conducted field tests.

\begin{equation}
Q=\begin{bmatrix}
0.001 & 0 & 0\\
0 & 0.0001 & 0\\
0 & 0 & 0.01
\end{bmatrix},\
R=\begin{bmatrix}
2.0 & 0 \\
0 & 0.001
\end{bmatrix}\label{eq4A1}
\end{equation}
\subsection{Configuration of Experimental System}
We validated the proposed attitude control algorithm using an actual cutting device mechanism of a Korean cabbage harvester. The cutting device is equipped with an IMU(In-ertial measurement unit. MPU-6050) sensor and a linear potentio-meter(KTC-100 mm), and the MCU(Micro control-ler unit, Arduino MEGA 2560) implements the sensor fusion part of Fig. \ref{f8}.

The control part maintains the control objective $\theta_{c}$ by calibrating $L_{\theta c}$, the 
\textcircled{\small 1} cylinder stroke in Fig. \ref{f7}(b), to $\theta$, which represents the pitch axis angle data. The expression for obtaining the calibrated data is equivalent to that of (\ref{calithata}), and the control objective $\theta_{c}$ is set to 35$^\circ$. In addition, the slide axis calibration for the linear potentiometer length $L_p$ represents the height $H_{p}$ data, as presented in (\ref{caliheight}). The \textcircled{\small 2} cylinder stroke $L_{xc}$ is controlled such that the control objective $H_c$ can always be maintained.
\begin{equation}
L_{\theta c}=-34.304\times\theta+1949.9\label{calithata}
\end{equation} 
\begin{equation}\label{caliheight}
H_{p}=1.0323\times L_{p}+16.177
\end{equation}
To quantitatively evaluate the performance of the attitude control system, we compared by computing the root mean square error (RMSE) by expression (\ref{RMSE}).
For RMSE, we referred to the presented in\citep*{chai2014root}.
\begin{equation}\label{RMSE}
RMSE = \sqrt{\frac{\sum_{i=1}^{n}(Reference_{i}-Actual_{i})^{2}}{n}}
\end{equation}
\begin{figure}
    \centering
    \begin{subfigure}[!t]{1.0\linewidth}
        \centering
        \includegraphics[width=\linewidth]{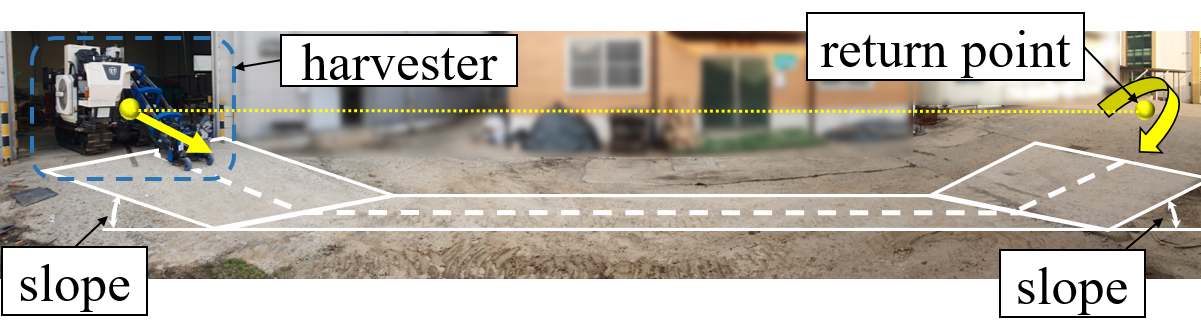}
        \caption{}
    \end{subfigure}
    \begin{subfigure}[!t]{1.0\linewidth}
        \centering
        \includegraphics[width=\linewidth]{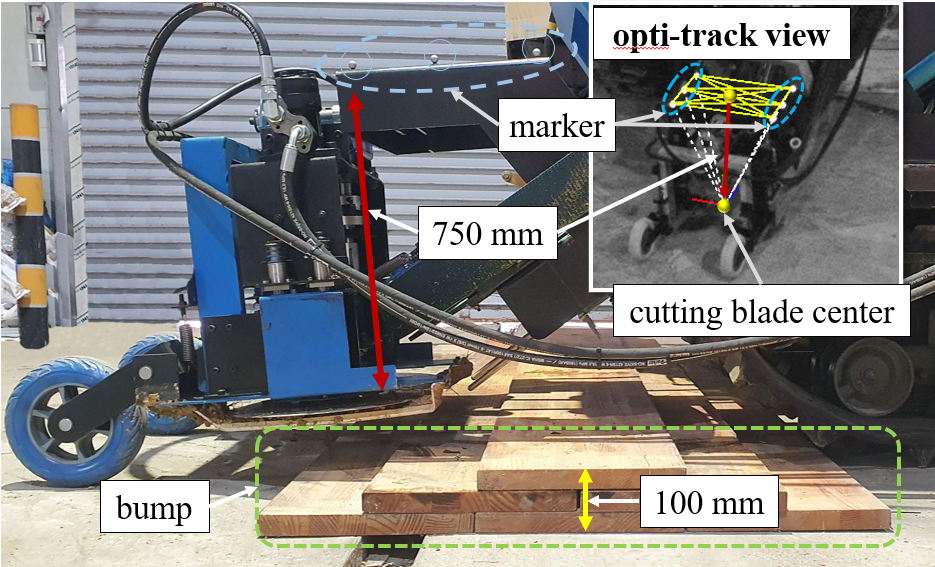}
        \caption{}
    \end{subfigure}
\caption{Experimental setup for attitude control system; (a) $\theta$ attitude control, (b) $H_{p}$ attitude control}
\label{f12}
\end{figure}
\begin{figure}
    \centering
    \begin{subfigure}[!t]{.9\linewidth}
        \centering
        \includegraphics[width=\linewidth]{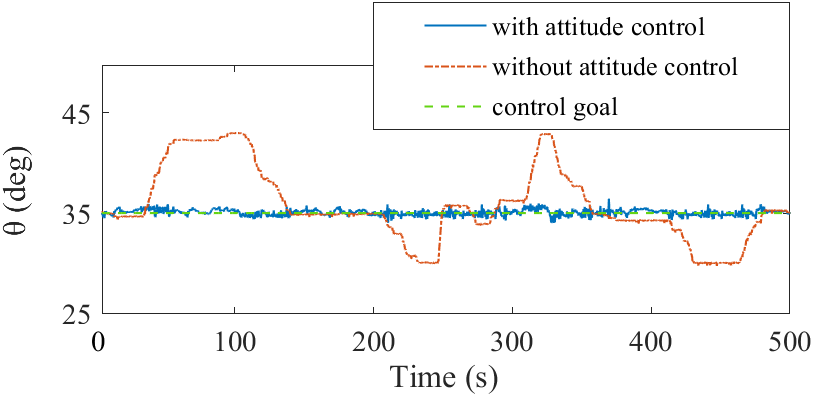}
        \caption{}
    \end{subfigure}
    \begin{subfigure}[!t]{.9\linewidth}
        \centering
        \includegraphics[width=\linewidth]{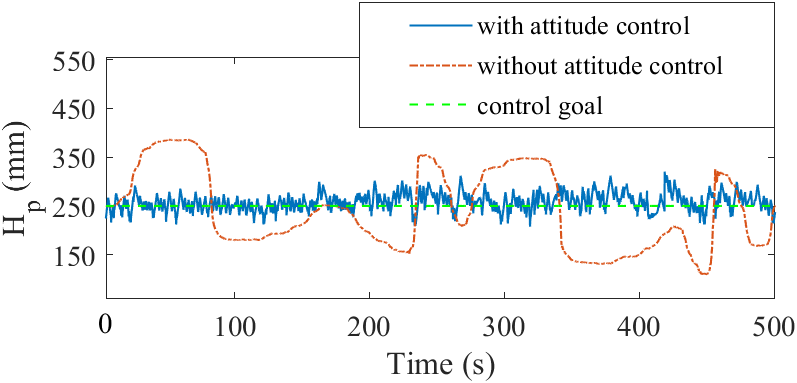}
        \caption{}
    \end{subfigure}
\caption{Comparison of cutter angle with / without attitude control system; (a) Compare $\theta$ controls, (b) Compare $H_{p}$ controls.}
\label{f15}
\end{figure}
\begin{figure*} 
    \centering
    \begin{subfigure}[b]{.78\linewidth}
        \centering
        \includegraphics[width=\linewidth]{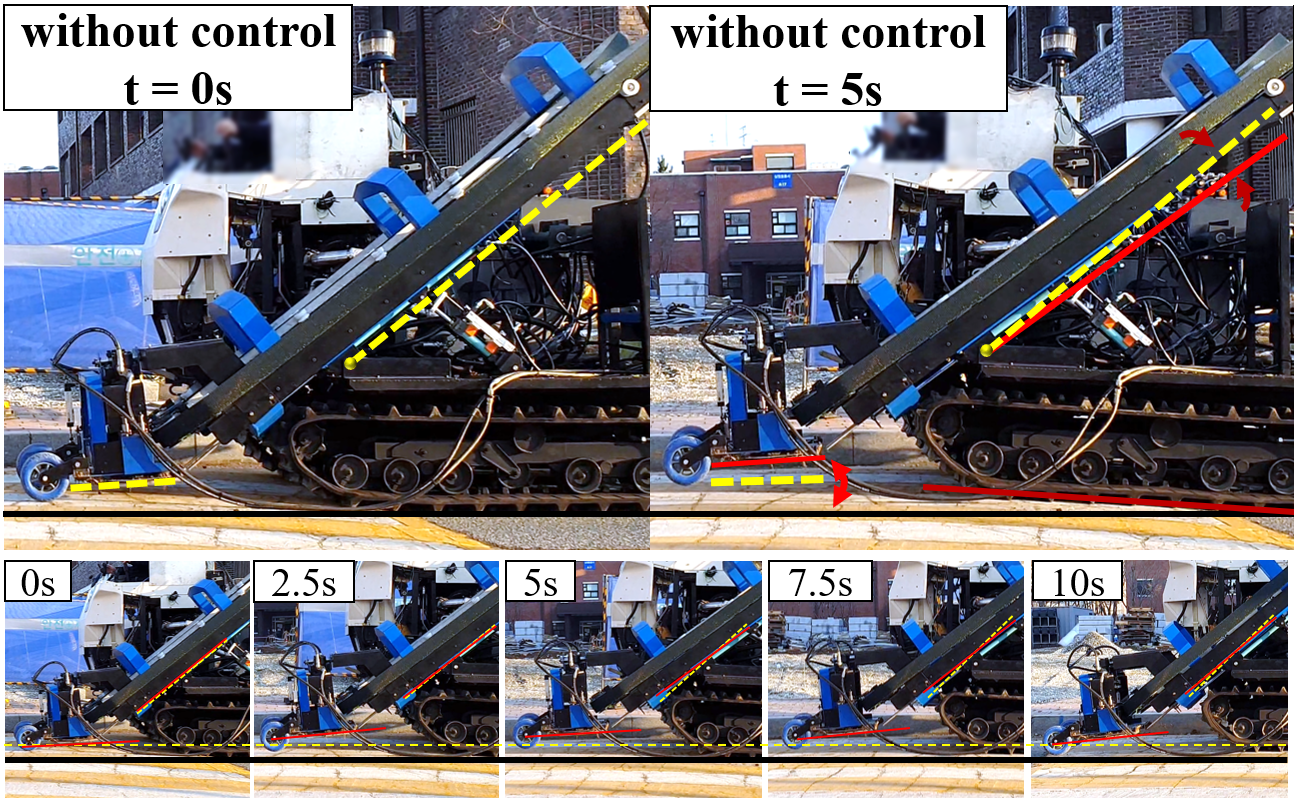}
        \caption{}
    \end{subfigure}
    \begin{subfigure}[b]{.78\linewidth}
        \centering
        \includegraphics[width=\linewidth]{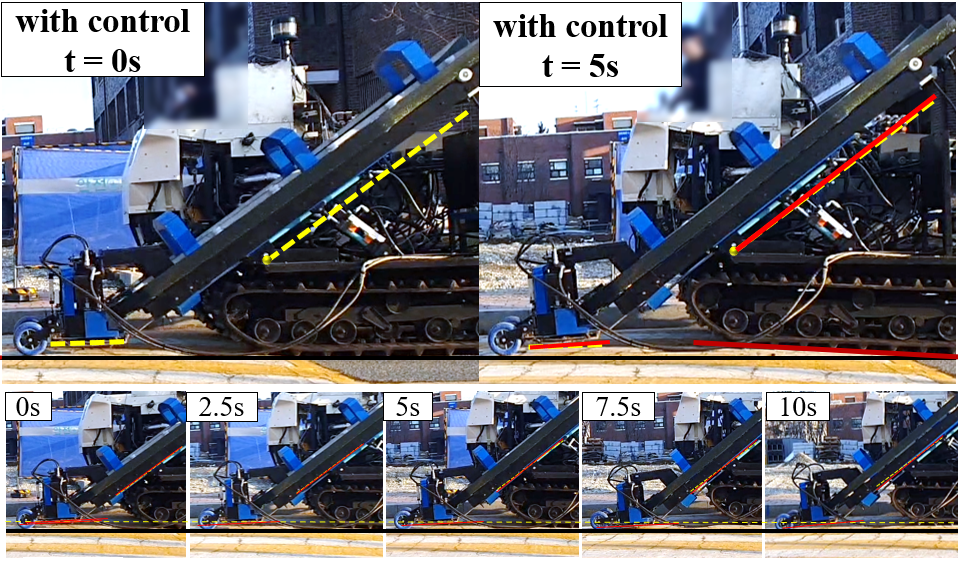}
        \caption{}
    \end{subfigure}
\caption{A snapshot of a cutting device of Korean cabbage harvester passing by an challenge was taken every 2.5 s. The yellow line is the reference line and the red line is the measured position line; (a) Without attitude control, (b) With attitude control.}
\label{f13}
\end{figure*} 
\begin{figure*} 
    \centering
    \begin{subfigure}[!t]{.475\linewidth}
        \centering
        \includegraphics[width=\linewidth]{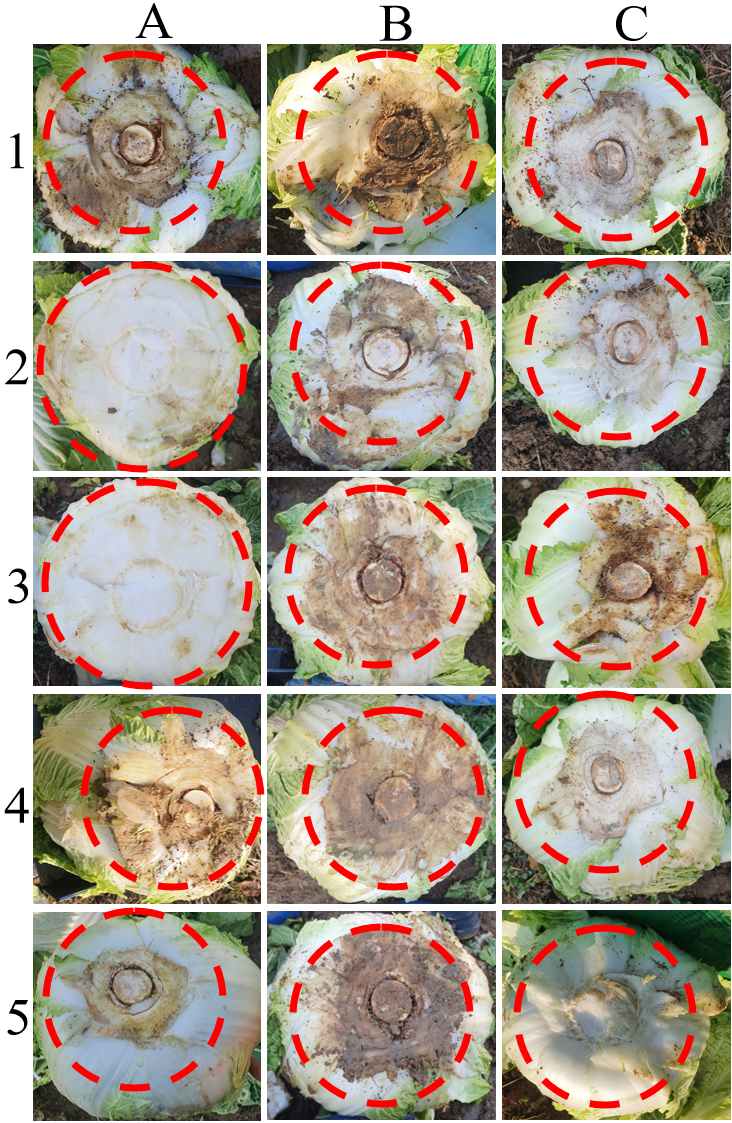}
        \caption{}
    \end{subfigure}
    \begin{subfigure}[!t]{.477\linewidth}
        \centering
        \includegraphics[width=\linewidth]{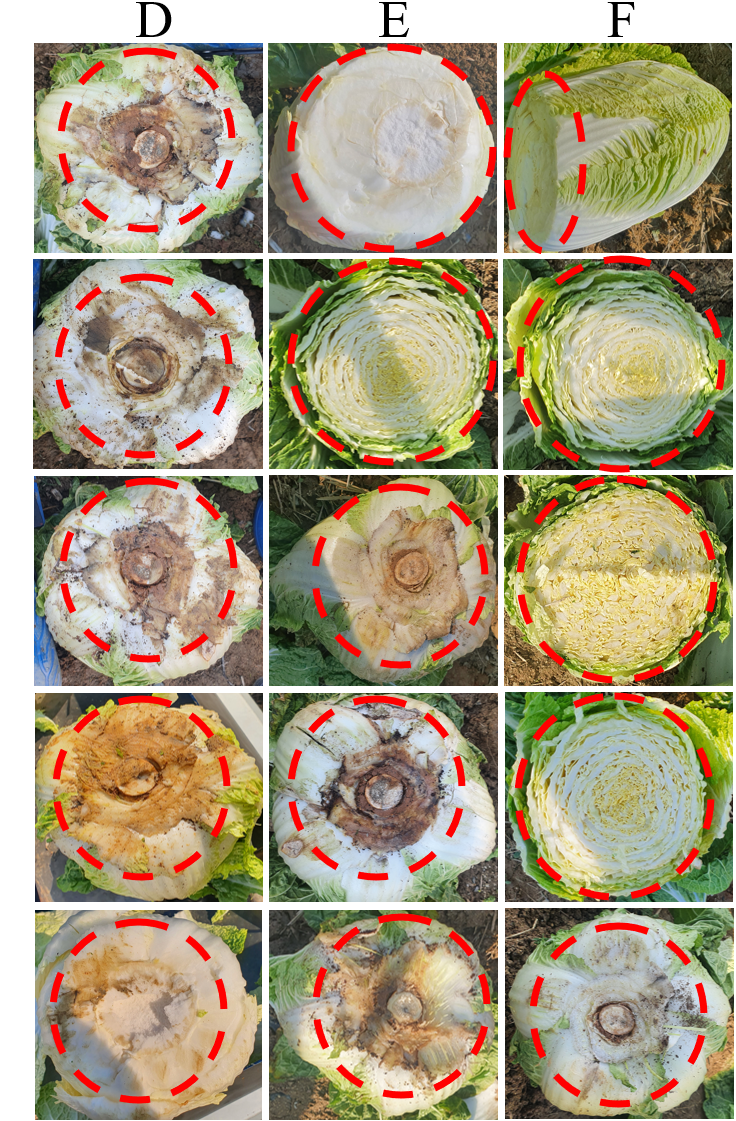}
        \caption{}
    \end{subfigure}
\caption{Comparison of cutting Korean cabbage with / without attitude control system; (a) Without attitude control, (b) With attitude control.}
\label{f16}
\end{figure*}
\subsubsection{$\theta$ attitude maintenance RMSE calculation}
For the RMSE comparison of pitch axes, the method for obtaining $Actual_{i}$ from (\ref{RMSE}) is set as $\theta$ and measured by IMU sensors of the $Actual_{i}$. Next, the method for obtaining $\theta$ without the attitude control system is presented as follows: Because we already calibrated the pitch axis angle $\theta$ in (\ref{calithata}) to the $L_{\theta c}$ in Fig. \ref{f7}(b), $\theta$ can be determined by $L_{\theta c}$. Therefore, the stroke of the cylinder operating during the postural control system driving indicates the extent to which the angle $\theta$ of the current cutting device is sloped. Based on this, we compare the $\theta$ angles obtained with and without the attitude control system. We then set $Reference_{i}$ to the height control objective $\theta_{c}$. The experiment was conducted by driving about 25 m through a straight course with an inclined plane as illustrated in the Fig. \ref{f12}a and returning. Furthermore, we obtained the RMSE for quantitative comparison.
\subsubsection{$H_{p}$ attitude maintenance RMSE calculation}
$H_{p}$ control can also obtain references with cylinder stro-kes, such as $\theta$, but it is impossible to measure when the guide wheels fall off the field. Therefore, the experimental system was constructed using eight motion capture cameras (Opti-Track Inc.), as illustrated in Fig. \ref{f12}b. In addition, six markers were attached to the lower part of the cutting device to verify the height retention performance of the attitude control system. When the marker is attached to the cutter blade, the camera is not recognized, so it is attached to the top of the cutting device and the marker point is adjusted below 750 mm, the distance to the blade, and set the length between the ground and the cutting blade center point as $Actual_{i}$ from  (\ref{RMSE}). We then set $Reference_{i}$ to the height control objective $H_{c}$. An arbitrary height of 100 mm bump was created, for 500 s, the experiment was drove out by repeating the forward and backward movements. Furthermore, we obtained the RMSE for quantitative comparison.
\subsection{Comparison of with/without Attitude Control System}
First, for the experiments, we set the control objective $\theta_{c}$ to 35$^\circ$. The results obtained from the experiments indicate that when the attitude control systems are absent, $\theta$ changes significantly depending on the slope angle, as illustrated in Fig. \ref{f15}(a). Conversely, with an attitude control system, it can be observed that the control objective of 35$^\circ$ can be maintained. The RMSE was able to confirm a 92$\%$ improvement from 3.66$^\circ$ to 0.29$^\circ$.

For the height retention experiments, we set the control objective $H_{c}$ to 250 mm. The reason for the high control target is that when passing through obstacles without using a attitude control system, the blades and guides of the cutting device are dragged to the ground, which exposes the device to damage. The results obtained from the experiments indicate that the height $H_{p}$ varies significantly in the absence of an attitude control system, as demonstrated in Fig. \ref{f15}(b). It can be observed from the graph that the control objective of 250 mm can be maintained. The RMSE was able to confirm a 77$\%$ improvement from 81.44 mm to 18.97 mm.

Fig. \ref{f13} demonstrates the difference between operations with or without the attitude control device. Therefore, we infer that the attitude control system of this study can enhance the cutting device and provide accurate cutting process.
\subsection{Experiment of Korean Cabbage Cutting}

As illustrated in Fig. \ref{f6}, for the experiment of cutting Korean cabbage, a field was constructed with a height of 100 mm and a distance between cabbages of 500 mm, similar to the actual source of the cabbage field. In addition, 15 cabbage cutting experiments were conducted when with and without attitude control. Korean cabbage was cut as illustrated in Fig. \ref{f16}. We tried to confirm the harvesting performance by checking whether the head was damaged by the cut surface and scoring accordingly. The harvest performance was determined as illustrated in Table. \ref{t2}. $\bigcirc$ are cut well, so 100 scores, $\square$ are so 80 scores, which can be the same as cut well if you take the steps to clean up the roots and outer leaves after transfer, $\triangle$ are so 50 scores because there was some damage to the heads due to over cutting or side cutting, and 0 scores to × because they were not worth a product. As a result, the score is 89 scores for with attitude control and 56 scores for without attitude control, indicating that improvements have been made due to attitude control system.
\begin{table}
\centering
\caption{RMSE of Pitch/Height with/without Attitude Control}
\begin{tabular}{c|ccc|ccc}
\noalign{\smallskip}\noalign{\smallskip}\hline\hline
\multirow{2}*{} & \multicolumn{3}{c|}{With attitude control} & \multicolumn{3}{c}{Without attitude control}\\
\cline{2-7}
   & \quad A & \quad B & C & \quad D &\quad  E & F \\\hline
 1 & \quad $\bigcirc$ & \quad $\square$ & $\bigcirc$ & \quad $\bigcirc$ & \quad $\triangle$ & ×  \\
 2 & \quad $\triangle$ & \quad $\bigcirc$ & $\bigcirc$ & \quad $\bigcirc$ & \quad × &  × \\
 3 & \quad $\triangle$ & \quad $\bigcirc$ & $\bigcirc$ & \quad $\bigcirc$ & \quad $\square$ & × \\
 4 & \quad $\square$ & \quad $\bigcirc$ & $\bigcirc$ & \quad $\square$ & \quad $\bigcirc$ & × \\
 5 & \quad $\bigcirc$ & \quad $\bigcirc$ & $\square$ & \quad $\triangle$ & \quad $\square$ & $\bigcirc$\\
\hline
Average& \multicolumn{3}{c|}{89} & \multicolumn{3}{c}{56}\\
\hline
\hline
\end{tabular}\\
$\bigcirc$:100, $\square$:80, $\triangle$:50, ×:0
\label{t2}
\end{table}
\section{Conclusions}
\label{sec:5}
In this study, we devised a mechanism for single- attitude control of cutting devices, not body attitude control, to solve the problem of driving platform body slope, which is a problem with conventional Korean cabbage harvesters, we designed a mechanism for the single-attitude control of cutting devices. In addition, we adopted the Kalman filter, which is widely used as a basic study for designing algorithms that control mechanisms. This was used to fuse the accelerometer, gyro sensors, and linear potentiometer with sensors. and, It was stabilized using a PID control. To verify the performance of this algorithm, we applied the algorithm to the attitude control mechanism. We then conducted a field test, and when applying the proposed attitude control algorithm through the results obtained from the experiments, the performance of $\theta$ and $H_{p}$ attitude control maintenance was improved by 92$\%$, 77$\%$ respectively. The performance of cutting Korean cabbage also improved from 56 scores to 89 scores. Furthermore, this improved performance obtained from this study is expected to exhibit better improvement results provided the angular and height changes, such as obstacles and thickness, are more significant than the experimental environment in this study. Through this study, we tried to suggest the future direction of the Korean cabbage harvester, and it is judged that it can be applied to other fields. In the future, we will study more accurate attitude control and cutting processes using a backstepping control for robust control.

\printcredits
\nolinenumbers
\bibliographystyle{model5-names}

\bibliography{main}


\bio{}
{Yonghyun Park} received the B.S. degree from the Department of Bio-indus-trial Machinery Engineering, Gyeongsang National University, Korea, in 2019. He is currently pursuing an M.S. degree in Biosystems Engineering from Gyeongsang National University. His research interests include field robotics, mechanical engineering, and harvesting robot.
\endbio

\bio{}
{Jongpyo Jun} received the B.S. degree in the Department of Mechanical Engineering from Sunchon National University in 2008 and M.S degree in the Department of Mechanical Engineering from Chonnam National University in 2015. Currrently, he is Ph.D. candidate in the Department of Rural and Biosystems Engineering from Chonnam National University. His research interests include field of mechanical engineering, design, and harvesting robot.
\endbio

\bio{}
{Hyoung Il Son} (M’11) received the B.S. and M.S. degrees from the Department of Mechanical Engineering, Pusan National University, Korea, in 1998 and 2000, respectively and the Ph.D. degree from the Department of Mechanical Engineering, KAIST (Korea Advanced Institute of Science and Technology), Korea in 2010.  In 2015, he joined the faculty of the Department of Rural and Biosystems Engineering, Chonnam National University, Gwangju, Korea, where he is currently an Associate Professor. Before joining Chonnam National University, from 2012 to 2015, he lead the Telerobotics Group, Central Re- search Institute, Samsung Heavy Industries, Daejeon, Korea as a Principal Researcher. He also had several appointments both academia and industry as a Senior Researcher with LG Electronics, Pyungtaek, Korea (2003–2005) and Samsung Electronics, Cheonan, Korea (2005–2009), a Research Associate with the Institute of Industrial Science, The University of Tokyo, Tokyo, Japan (2010), and a Research Scientist with Max Planck Institute for Biological Cybernetics, Tübingen, Germany (2010–2012). His research interests include field robotics, agricultural robotics, haptics, teleoperation, and discrete event and hybrid systems.
\endbio

\end{document}